\documentclass[twocolumn, 10pt]{article}
\usepackage[cm]{fullpage}

\usepackage{amssymb,amsfonts,times,url}
\usepackage{graphicx}
\usepackage[round]{natbib}

\begin{document}
\title{Pairwise alignment incorporating dipeptide covariation.}
\author{Gavin E. Crooks\footnote{To whom correspondence should be addressed}~\footnotemark[2]\  , Richard E. Green\footnote{These authors contributed equally to this work.}~  and  Steven E. Brenner
\\ Dept. of Plant and Microbial Biology, \\111 Koshland Hall \#3102, \\University of California,
Berkeley, \\ CA 94720-3102, USA}

\maketitle

\begin{abstract}
%
{\bf Motivation:}
Standard algorithms for pairwise protein sequence alignment make the simplifying assumption that amino acid substitutions at neighboring sites are uncorrelated. This assumption allows implementation of fast algorithms for pairwise sequence alignment, but it ignores information that could conceivably increase the power of remote homolog detection.  We examine the validity of this assumption by constructing extended substitution matrixes that encapsulate the observed correlations between neighboring sites, by developing an efficient and rigorous algorithm for pairwise protein sequence alignment that incorporates these local substitution correlations, and by assessing the ability of this algorithm to detect remote homologies.
%
\\ {\bf Results:}
Our analysis indicates that local correlations between substitutions are not strong on the average. Furthermore, incorporating local substitution correlations into pairwise alignment
did not lead to a statistically significant improvement in remote homology detection. Therefore, the standard assumption that individual residues within protein sequences evolve independently of neighboring positions appears to be an efficient and appropriate approximation.
%
\\ {\bf Availability:}
Sequence data, software, and matrixes are freely available from 
{\tt http://compbio.berkeley.edu/}.
%
\end{abstract}

\section{INTRODUCTION}

Among the most commonly used tools in computational biology are the pairwise protein sequence alignment methods, such as SSEARCH, FASTA and BLAST  \citep{SmithWaterman1981,Pearson1988,Altschul1990,DurbinEddy1998}.
These algorithms are elegant, efficient and effective methods of detecting similarity between closely related protein sequences.  However, the ability of fast pairwise methods to detect homology deteriorates as the divergence between the sequences increases.  Past the ``twilight zone'' (20-30\% pairwise sequence identity), only a small fraction of related proteins can be found.  \citep{Sander1991,Doolittle-1992-ProtSci,Brenner-1998-PNAS,Green-2002-ProcIEEE}.  Therefore, in order to make better use of the vast and increasing amount of available biological sequence data, there is an immediate need for more sensitive, fast database search methods.

For the sake of computational efficacy, current pairwise alignment methods make several simplifying assumptions. First, amino acid substitutions are assumed to be homogeneous between protein families.  The most commonly used substitution matrices (BLOSUM \citep{HenikoffHenikoff-1992-PNAS} and PAM \citep{Dayhoff1978}) are thus generic models of protein sequence evolution across all protein sequence families at various evolutionary distances.  Second, substitutions at a given site are assumed to be uncorrelated with those on neighboring sites.  That is, the likelihood of substituting an amino acid X for amino acid Y is assumed to be independent of the sequence context of X.  It is known that both of these simplifying assumptions introduce errors into homology searching.  Relaxing the assumption of homogeneous substitution across protein families can significantly improve the performance of pairwise alignment methods  \citep{Yu-2003-PNAS}. Furthermore, alignment methods that remove the assumption of homogeneity among different positions in the sequence, and instead model the heterogeneity of the given protein family, have been found to be dramatically superior for remote homology detection \citep[R. E. Green and S. E. Brenner, Unpublished data]{Park-1998-JMB}.  Unfortunately, these profile methods (e.g. PSI-BLAST \citep{Altschul1997}, HMMER \citep{HMMER}, SAM \citep{SAMT98} etc.~) are not tractable for all query sequences.  They require the presence, identification, and correct alignment of homologous sequences in order to generate a model of the query sequence's family.  Therefore, the fast and universally applicable pairwise methods remain widely used for database searching, despite their lower sensitivity.

One proposed strategy for increasing the sensitivity of pairwise alignment is to use a more sophisticated scoring function for amino acid substitutions, namely one that is sensitive to the sequence context in which the residue resides.  For example, amino acid sequences are correlated with secondary structural features, such as helixes and loops, which can directly lead to structure  dependent substitution patterns \citep{Thorne1996,Topham1997,Goldman1998}.  Similarly, one might intuitively expect structurally and functionally important residues, such as cysteines and prolines,  to be more or less conserved depending on their local sequence environment and the prevalence of particular motifs.

The first large-scale exploration of the effect of sequence context on amino acid evolution
was performed by Gonnet and co-workers \citep{Gonnet-1994-BiochemBiophys}, who examined the frequencies of dipeptide substitutions, and compared them to the dipeptide substitution frequencies expected assuming no sequence dependent correlations.  Despite the fact that nearly half of the elements of the $400 \times400$ observed dipeptide matrix were vacant (due to the sparsity of data) several interesting patterns were evident.  
The chief trend was that amino acids are generally more likely to be conserved if they are adjacent to positions that are also conserved.

More recently  \citet{JungLee-2000-ProteinScience} have taken advantage of the large increase in available data to reexamine trends in dipeptide evolution.  They used the observed patterns of substitution within a large set of structure-based alignments to generate dipeptide substitution matrices.  Furthermore, they developed an extension to the standard Smith-Waterman alignment algorithm that incorporates a term from these dipeptide matrices.  By using sequence and structure context information, they show some improvement in homolog detection in a limited test set.  However, their method could not be extensively tested or practically utilized because an efficient dynamic programming method for finding the optimal alignment was not known to the authors.  Instead, they adopted a heuristic search that is not guaranteed to find optimal alignments.

In this study, we have extended the work described above by examining the strength of local, dipeptide substitution correlations using the massive amount of alignment data within the BLOCKS database.  We have also extended the standard Smith-Waterman algorithm to include local dipeptide correlation information over a user-defined distance.  Like Smith-Waterman, this new polynomial time algorithm, {\tt doublet}, finds the optimal alignment under the scoring scheme described.  Using a standard remote homolog detection evaluation strategy, we have tested {\tt doublet} against the Smith-Waterman algorithm to measure the impact of including this extra information.  Perhaps surprisingly, we found that incorporating doublet substitution correlations leads to a statistically insignificant difference in homology detection.


\section{Methods}

\subsection{Quantifying substitution correlations}

Consider two aligned, ungapped sequences, ${\mathrm x}=x_1,x_2,\cdots,x_n$ and ${\mathrm y} = y_1,y_2,\cdots,y_n$, both of length $n$, where each element  represents one of the 20 canonical amino acids. We wish to use the patterns of conservation and variation between these sequences to estimate the log odds that the sequences are homologous (i.e., that both sequences have descended from a common ancestor). 
\begin{equation}
{\mathcal S} 
= \log \frac{ q(\mathrm{x} ; \mathrm{y} )} {p(\mathrm{x} ) p(\mathrm{y} ) } 
= \log 
\frac{q(x_1, x_2, \cdots, x_n; y_1, y_2, \cdots, y_n) 
}{ p(x_{1}, x_{2},  \cdots, x_{n} ) p(y_{1}, y_{2},   \cdots, y_{n} )  }
\, .
\label{similarity}
\end{equation}
\noindent Here,  $p( \mathrm{x})$ is the background probability of the given amino acid segment and $q( \mathrm{x} ; \mathrm{y})$ is the target probability of observing the pair of segments in diverged homologous sequences.

Except for very short segments, the background and target probability distributions are large and cannot be directly measured.  Therefore, Eq.~\ref{similarity} is typically simplified by assuming that substitutions probabilities are homogeneous (independent of the location in the fragment) and that both the substitutions and the sequences themselves are uncorrelated from one position to the next.
Consequentially, the total similarity score is now a sum of independent parts, 
\begin{equation}
{\mathcal S} \approx \sum_k s(x_k ;y_k), \quad 
s(i; j ) = \log \frac{q(i; j )} {p(i) p(j)}
\label{singlet}
\end{equation}
\noindent
The log odds of residue replacement, $s(i, j )$, is an element of a standard singlet substitution matrix, of the type widely used in pairwise sequence alignment  \citep{Altschul-1991-JMB}.

This approximation of the full similarity by a sum of singlet substitution scores requires that we neglect all inter-site correlations. We can perform a more controlled approximation by noting that a homogeneous multivariate probability can be expanded into a product of single component distributions, pairwise correlations, triplets correlations, and so on. 
\begin{eqnarray}
\lefteqn{  
                 P(z_1, z_2,\cdots, z_n) =  \prod_i P( z_i) \times 
                 \prod_{i<j}\frac{P(z_i, z_j)}{ P(z_i) P(z_j)} 
} \nonumber
\\ 
& & \times \prod_{i<j<k}\frac{P(z_i, z_j, z_k) P(z_i) P(z_k) P(z_j)}{ P(z_i,z_j) P(z_i, z_k) P(z_j, z_k)} 
\ldots
\end{eqnarray}
\noindent If we assume that substitution probabilities are independent of the location within the fragment, then we can apply this expansion to the segment homology score (Eq.~\ref{similarity}). 
\begin{equation}
{\mathcal S}  =\sum_{k=1}^n s(x_k; y_k) 
+\sum_{l=1}^{L}  \sum_{k=1}^{n-L} d_{l}(x_k, x_{k+l};y_k, y_{k+l}) 
+ \ldots
\label{expansion}
\end{equation}
The first term of this expansion represents single residue replacements, as in Eq.~\ref{singlet}.
The next term defines the doublet substitution scores,
\begin{equation}
d_{l}(i,i'; j,j')
= \log \frac{q_l(i,i';j,j')} {p_l(i,i') p_l(j,j')} - s(i; j) - s(i'; j')
\label{doublet_matrix}
\end{equation}
Here, $i$ and $i'$ are residues separated by a distance $l$ along one amino acid chain, while $j$ and $j'$ are the corresponding aligned residues on the putative homologous sequence; $q_l(i,i';j,j')$ is the target probability of observing this aligned quartet, and $p_l(i,i')$ is the background probability of this residue pair in protein sequences.  
These doublet scores represent the additional similarity due to correlations between substitutions.  

By truncating the expansion of the full similarity score at doublet terms (Eq.~\ref{expansion}), we are assuming that triplet and higher order correlations between substitutions are relatively uninformative. For reasons discussed below, this is probably a reasonable approximation. Furthermore, the most important inter-site correlations are between residues neighboring on the chain (Fig.~\ref{fig-MI}).  Therefore, we can restrict the maximum distance over which doublet interactions are scored without serious error.

The average similarity score is the inter-homolog mutual information $I$ \citep{CoverThomas},
a measure of the inter-sequence correlations. A high mutual information value indicates strong correlation, whereas a mutual information value of zero indicates uncorrelated variables. Mutual information has various advantages as a correlation measure: it is firmly grounded in information theory, it is additive for independent contributions and it has consistent, intuitive units (bits).
\begin{equation}
I(\mathrm{x};\mathrm{y}) = \sum q(\mathrm{x},\mathrm{y}) \log_2 \frac{q(\mathrm{x},\mathrm{y})}{p(\mathrm{x})p(\mathrm{y})}	
\label{eq-mi}
\end{equation}
The average singlet score is the inter-homolog mutual information per residue, under the assumption that replacements are uncorrelated. This is frequently reported as the  ``relative entropy'' of the substitution matrix. The average doublet score is the first order correction to the inter-sequence mutual-information due to inter-site correlations.
Consequentially, we may evaluate the comparative importance of singlet and doublet contributions to the sequence similarity by examining the average contributions of these different components to the full inter-homolog mutual information. 

The preceding analysis applies to contiguously aligned sequence segments. However, in addition to substitutions, protein sequences are modified by the insertion and deletion of residues. Since it is not obvious how to capture the existence of indels in doublet scores, in the following discussion we assume that dipeptide correlations do not extend across gaps, and we adopt the simple and standard affine model of gap lengths.   This approximation should have little impact, since aligned detectably homologous sequences tend to have relatively few indels, particulary in regions that are significantly similar.

\subsection{Alignment Algorithm}

\begin{figure}[t!]
\begin{center}
\includegraphics[width=3.25in]{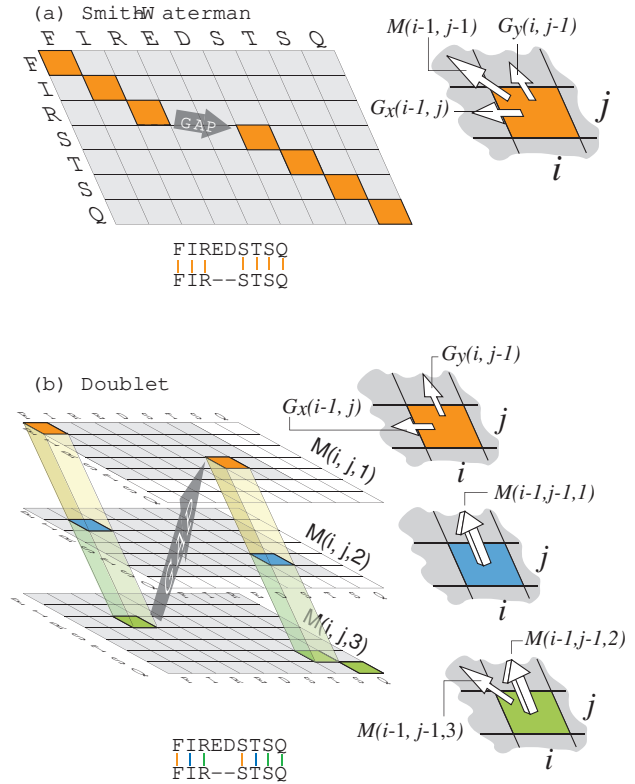}
\caption{A comparison of Smith-Waterman and {\tt doublet} 
sequence alignment. (a) A Smith-Waterman match table, with the optimal alignment highlighted. The value of each cell is the maximum of 1. the singlet match score  (this is the start of an alignment ), 2. the singlet score plus the match score from the previous cell along the diagonal (this extends an aligned region), or 3. the singlet score plus the optimal score from a gap score table (the previous residue was not aligned)
(b) For {\tt doublet}, multiple match tables are used (Eqs.~\ref{match1}-\ref{gaps}). The number of match tables is 1 plus the distance over which dipeptide correlation information is considered (in this example, 2). Again, the optimal alignment is highlighted. The top table corresponds to the starts of aligned regions, the middle table corresponds to aligned regions of at least 2 consecutive residues and the bottom table corresponds aligned regions of at least 3 consecutive residues. The alignment path through these tables falls through to lower tables in regions of conecutive aligned residues and begins again in the top table following gaps. To extend dipeptide context scoring over longer distances requires additional match tables.
}
\label{alignmentalgorithm}
\end{center}
\end{figure}

We have extended the standard Smith-Waterman optimal local sequence
alignment algorithm  \citep{SmithWaterman1981} to incorporate doublet
substitution scores (See Fig.~\ref{alignmentalgorithm}).  The time complexity of Smith-Waterman is $O(nm)$,
where $n$ and $m$ are the lengths of the two sequences.  Adding
doublet scores increases the complexity to $O(nmL)$, where
$L$ is the distance over which substitution correlations
are scored.  This efficient dynamic programming alignment is possible because, although we are scoring correlations between residues that are not directly aligned, these correlations are local along the chain. The space complexity of our implementation is also $O(nmL)$; this could be improved using standard techniques \citep{DurbinEddy1998}.

The additional similarity score associated with adding the final match pair  $x_i, y_j$ to the alignment contains singlet ($S$) doublet ($D$) substitution scores;
\begin{eqnarray}
	S(i,j) &=& s(x_i, y_j)  , \\
D(i,j,r) &=& \sum_{l=1}^{r} d_{l}(x_{i-l}, x_i; y_{j-l}, y_j)  .
\end{eqnarray}
\noindent 
Here, $r$ is the length of the preceding contiguous segment of aligned residues, or the maximum sequence separation over which doublet correlations are scored, whichever is less. Deletions of length $k$ are weighted with the affine penalty $ -(g_{\mathrm{open}} + (k-1) g_{\mathrm{ext}})$, where $g_{\mathrm{open}}$ and $g_{\mathrm{ext}}$ are positive constants. This standard affine gap length model is both computationally efficient and surprisingly effective.  \citep{SmithWaterman1981,Altschul1986,Zachariah2005}.

The optimal, highest scoring alignment between two sequences ($\mathrm{x} = x_1,x_2,\cdots, x_n$ and $\mathrm{y} = y_1,y_2,\cdots, y_m$) is found by populating a series of score tables, also known as dynamic programming matrices. 
The entries of the match table, $M(i,j,r)$, are the maximum alignment score for an alignment that terminates with an ungapped segment of length $r$, ending at the $i$th position of $\mathrm{x}$, and the $j$th position of $\mathrm{y}$. Similarly, the gap tables $G_x(i,j)$ and $G_y(i,j)$ contain the maximum alignment similarity given that the alignment ends with $x_i$ or $y_j$ gapped.
The entries of these tables can be efficiently computed starting from the following boundary conditions:
$M(i,0,l), M(0,j,l), G_{x/y}(i,0),  G_{x/y}(0,j) = -\infty$.
%
A single aligned amino acid pair may signal the beginning of a new local alignment, or it may occur immediately after any alignment gap. 
\begin{equation}
M(i,j,1) = \max  \left\{
    \begin{array}{l}
       S(i,j)\\
      S(i,j) + G_x(i-1, j) \\
      S(i,j) + G_y(i, j-1)  \\      
    \end{array} 
    \right.
    \label{match1}
\end{equation}
In standard Smith-Waterman this is the only necessary match score table. However, in {\tt doublet} we require additional match tables so that we may keep track of match scores over extended, contiguously aligned regions. Of necessity, longer ungapped segments occur only after shorter segments. We restrict the maximum distance $L$ over which doublet correlations are scored, since
we expect that the useful information that can be extracted from doublet correlations will decay rapidly with sequence separation (See Fig.~\ref{fig-MI}).
Consequentially, we do not need to explicitly consider ungapped segments of length greater than $L+1$.%
\begin{eqnarray}
M(i,j,2\leq r \leq L)  &=&
       S(i,j) + D(i,j,r-1)  \\ 
      && \quad {}+ M( i-1, j-1, r-1)  \nonumber
 \\    
M(i,j,L+1) &=& S(i,j) + D(i,j,L)  \nonumber\\
&& \!\!\!{}+ \max  \left\{ 
    \begin{array}{l}
      M( i-1, j-1, L)  \\
      M( i-1, j-1, L+1) 
    \end{array} 
    \right.       \nonumber
\end{eqnarray}
Gaps in the alignment are either preceded by a match or they extend an existing gap.
\begin{eqnarray}
G_x(i,j) &=& \max_{r=1,L}  \left\{
    \begin{array}{l}
      M( i-1, j-1, r) -  g_{\mathrm{open}}\\ 
      G_x(i-1,j) -  g_{\mathrm{ext}}\\ 
    \end{array} 
    \right.
     \nonumber \\
G_y(i,j) &=& \max_{r=1,L}  \left\{
    \begin{array}{l}
      M( i-1, j-1, r) - g_{\mathrm{open}}\\ 
      G_y(i,j-1) -  g_{\mathrm{ext}}\\ 
    \end{array} 
    \right.    
    \label{gaps}
\end{eqnarray}

The largest score within the match table marks the last aligned position of the optimal alignment. The full alignment can be found by backtracking through the table, according to the choices previously made during the scoring step.

We used the method of \citet{BaileyGribskov-2002-JCompBiol} to fit an extreme value distribution to the results of aligning a query sequence against a database of possible homologs. The maximum likelihood parameters are then used to assign E-values to each alignment.

\begin{figure*}[t]
\begin{center}
\includegraphics{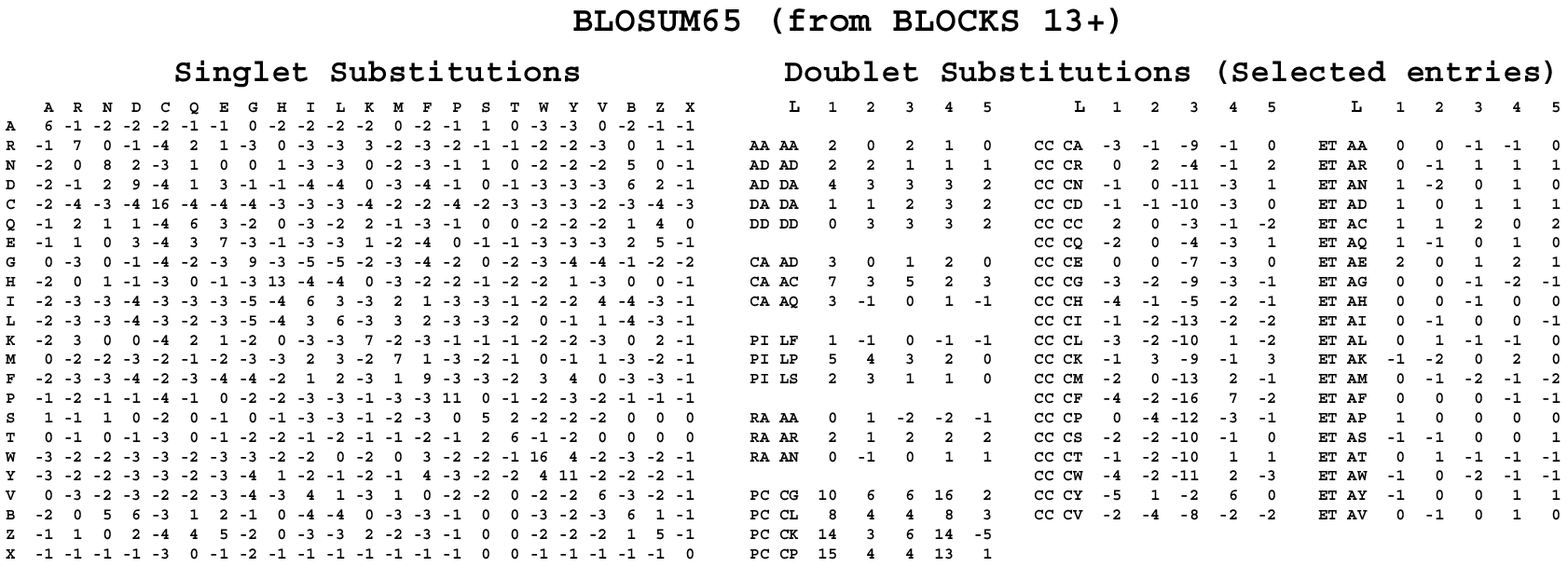}
\caption{BLOSUM65 singlet substitution matrix derived from the BLOCKS 13+ database (left), and selected elements of the corresponding doublet substitution matrices (right). Scores are in 1/4 bit units, rounded to the nearest integer.  The average standard statistical error is about 1/4 bits (i.e. about 1 unit) for the doublet scores, and essentially insignificant for the singlet scores, as judged by bootstrap resampling (See Sec.~\ref{sec-blosum}) The singlet scores are the log odds of observing the given substitution; positive scores are more likely, and negative score less likely to be observed than would be expected for uncorrelated sequences (Eq.~\ref{singlet}). Similarly, the doublet scores represent the log odds for observing pairs of substitutions, at various sequence separations, relative to the singlet substitutions likelihood  (Eq.~\ref{doublet_matrix}). For example, the L=3 column for ET AV (bottom right) indicates a score of zero for the alignment of ExxT in one sequence to AxxV in the other.
%
%
}
\label{fig-blosum}
\end{center}
\end{figure*}

\subsection{Doublet BLOcks SUbstitution Matrix}
\label{sec-blosum}

A doublet substitution matrix (Eq.~\ref{doublet_matrix}) contains $20^4 = 160,000$ entries, of which $20^2 \times(20^2+1)/2=80,200$ are unique due to the underlying symmetry, $d_l(i,i';j,j') = d_l(j,j';i,i')$. 
To accurately estimate these scores we require a very large collection of reliably aligned protein 
sequences.  The BLOCKS database is one such resource \citep{HenikoffHenikoff-1992-PNAS,Henikoff-2000-NAR-Blocks}. 
Each database block consists of a reasonably reliable, ungapped multiple sequence alignment of a core protein region. BLOCKS version 13+ contains $11,853$ blocks, containing, on average, 56 segments of average length 26 residues. Overall, about $10^9$ pairwise amino acid comparisons are available for study.

The widely used canonical BLOcks SUbstitution Matrixes (BLOSUM) were generated from version 5 of the BLOCKS database  \citep{HenikoffHenikoff-1992-PNAS}. In order to generate a series of matrices representing different evolutionary divergences, the sequences in each block are clustered at a given level of sequence identity and the inter-cluster sequence correlations are collected. Thus BLOSUM100 (where only 100\% identical sequences are clustered) represents a wide range, including low levels, of evolutionary divergence, whereas BLOSUM30 represents only correlations between very diverged sequences.

In principle, we should match the divergence inherent in the substitution matrix to the divergence of the pair of sequences we wish to align \citep{Bishop1986,Thorne1991,Thorne1992,Altschul1993}. However, this is computationally expensive, and, in practice, a single matrix is chosen based on its ability to align remote homologs, on the grounds that matching close homologs is relatively easy  \citep{Brenner1996,Brenner-1998-PNAS,Crooks2005a}. In a recent evaluation of remote pairwise homology detection efficacy \citep{Green-2002-ProcIEEE, Zachariah2005}, we discovered that the BLOSUM65 substitution matrix, reparameterized from the BLOCKS 13+ database, was more effective than any other reparameterized BLOSUM (BLOCKS 13+), classic BLOSUM (BLOCKS 5) or PAM  \citep{Dayhoff1978} substitution matrix, and was comparable to the most effective VTML matrix  \citep{Muller-2002-MolBiolEvol}. Consequentially, we have built  singlet and doublet substitution matrices from the BLOCKS 13+ database at 65\% clustering, using an adaptation of the original BLOSUM clustering code~\citep{HenikoffHenikoff-1992-PNAS}. This provides approximately  $10^7$ - $10^8$ independent aligned doublets, depending on the sequence separation $l$. 

The estimated doublet target frequencies $q_l(i,i' ; j,j')$ where smoothed and regularized by adding a pseudocount  $\alpha(i,i';j,j')$ to the raw count data, $n(i,j';j,j')$. These pseudocounts are taken to be proportional to the marginal singlet target probabilities, $q_l(i;j)q_l(i',j')$. %
\begin{eqnarray}
q_l(i,i';j,j')  &\approx&
\frac{ \alpha(i,i';j,j') + n(i,i';j,j')} {A+N}
\label{doublet_pseudocount1}
\\
\alpha(i,i';j,j') &=& A \times q(i;j) q(i';j')
\label{doublet_pseudocount2}
\end{eqnarray}
\noindent 
Where, $N$ is the total number of counts. Thus, if no data are available (the total number of counts is zero, $N=0$), then all doublet scores would be zero, as can be seen from  Eq.~\ref{doublet_matrix}.
Here, $A$ is a scale parameter that determines how much data is required to overcome the prior probability inherent in the pseudocount. Typically, such scale factors are picked empirically. However, in this case, we performed a full Bayesian analysis and determined that for doublet substitutions reasonable values of $A$ are about $2\times10^6$, which can be compared to the $10^7$ to $10^8$ actual observations. The full details are given in the supplemental materials. A representative subset of a doublet substitution matrix is shown in fig.~\ref{fig-blosum}.

Standard statistical errors were estimated by non-parametric Bayesian bootstrap resampling on sequence blocks  \citep{Efron1979,Rubin-1981-AnnStat}.
Instead of assigning equal weight to every sequence block, each block is instead given a random weight drawn from a Dirichlet distribution. This random reweighting induces random changes in the estimated scores, thereby providing an estimate of the statistical errors caused by the finite size and inhomogeneity of the training data.

\subsection{Evaluation of remote homology detection}

We have previously developed and applied a sensitive strategy for evaluation of
database search
methods \citep{Brenner-1998-PNAS,Green-2002-ProcIEEE,Zachariah2005,Price2005}. 
This 
strategy is made possible by the availability of a large collection of
protein sequences whose evolutionary interrelations are known (primarily from
structural information).
In our approach, each sequence is aligned against every other sequence, and the
alignment scores are used to determine putative homologs. We then
consider the proportion of correctly identified homologs as a function of
erroneous matches. 
 Because the homology information derives from
sequence-independent data, we avoid the circularity inherent
in other evaluation approaches.

The collection of related sequences is derived from the Structural
Classification Of Proteins (SCOP) database
\citep{Murzin-1995-JMB}.
We use the ASTRAL compendium
 \citep{Chandonia-2004-NAR}
of representative subsets of SCOP release 1.61 (Sept. 2002), filtered so that no two domains
share more than 40\% sequence identity.  We partition every other SCOP fold into separate test and training
subsets of approximately equal size, each containing about 550 superfamilies, 2500
sequences, and 50,000 homologous sequence pairs. 
To avoid over-fitting,
adjustable parameters are optimized using the training set.
Results
of an all-versus-all comparison of the test set, using these optimized
parameters, are reported as a plot of coverage (fraction of true
relations found) versus errors per query (EPQ), the total number of
false relations divided by the number of sequences (See Fig.~\ref{CVE}). 
The raw, unnormalized coverage is the fraction of all true relations
that are found.

Since the number of relations within a superfamily scales as the
square of the size of the superfamily, and because SCOP superfamilies
vary greatly in size, this reported coverage is dominated by the
ability to detect relations within the largest superfamilies. To
compensate for this unwarranted dependence, we also report 
 the average fraction of true
relations per sequence (linear normalization) and the average fraction
of true relations per superfamily (quadratic normalization). In
general, large superfamilies are more diverse, and the relationships
within them are harder to discover  \citep{Green-2002-ProcIEEE}. Thus,
unnormalized coverage is typically less than the linearly normalized 
coverage, which in turn is less than quadratically normalized
coverage. One important point of comparison for search results is 0.01
errors per query rate for linearly normalized results, the average fraction of true
relations per database query at a false positive rate of 1~in~100.
We
report the observed difference in coverage of two methods at this selected EPQ, and determine 
standard statistical errors and confidence
intervals using Bayesian bootstrap resampling 
\citep{Rubin-1981-AnnStat,Price2005}.

\section{RESULTS}

\begin{figure}[t]
\begin{center}
\includegraphics{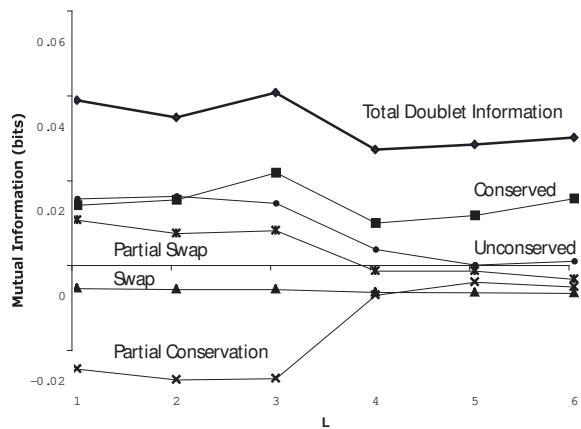}
\caption{The inter-sequence mutual information of homologs encoded in inter-site correlations at increasing separation, L. In other words, the average doublet substitution scores (Eq.~\ref{eq-mi}). The top, dark line is the total information at various sequence separations. For comparison, the information encoded in the corresponding singlet substitutions (the average singlet matrix score) is 0.31 bits per residue. The remaining lines illustrate the relative contributions of different substitutions classes to this total information; these are exact conservation XY$\leftrightarrow$XY, partial conservation XY$\leftrightarrow$XZ, swaps XY$\leftrightarrow$YX, partial swaps XY$\leftrightarrow$ZX, and unconserved, double substitutions XY$\leftrightarrow$ZU.
}
\label{fig-MI}
\end{center}
\end{figure}


\subsection{Doublet Substitution Correlations.}

Various trends are evident within the doublet score matrix, as illustrated in fig~\ref{fig-blosum}. Notably, exact conservations, such as AA$\leftrightarrow$AA, AD$\leftrightarrow$AD and DD$\leftrightarrow$DD, etc., generally have positive scores. 
This is expected because the pairs of sequences used to build the BLOSUM matrices have a variety of inter-sequence similarity, ranging from mostly conserved to very diverged. 
Thus the observation of a conserved residue suggests that the sequences are relatively undiverged, and therefore that other aligned  residues are also more likely than average to be conserved.

Also notable is that many (but far from all) exact swaps, such as DA$\leftrightarrow$AD, are significantly more likely that expected. Possibly, this is because the effect of a deleterious mutation X$\rightarrow$Y can sometimes be ameliorated by the occurrence of the corresponding mutation Y$\rightarrow$X, in the immediate sequence neighborhood. Partial swaps, where only one of the substitution pair is conserved, are also often positive. This might reflect alignment errors in the original dataset. The most highly positive scores (and therefore those events that are most over-represented in the data relative to uncorrelated substitutions) are associated with the substitutions PC$\leftrightarrow$Cx, i.e., a translocation of a cystine, replacing a proline.  The most relatively uncommon substitutions involve the mutation of one cystine in the cystine pair CxxC (second column), a widespread and important motif found, for example, in the thioredoxin
family. However, these interesting particular cases are atypical. Most of the doublet substitution matrix is similar to the ET$\leftrightarrow$Ax substitutions displayed in the third column; the majority of the scores are not significantly different from zero, indicating that most possible substitution doublets are essentially uncorrelated.

We can place the above observations on a quantitative footing by considering the
inter-sequence mutual information (Eq.~\ref{eq-mi}), a measure of the correlation strength between aligned homologous sequences. The first order contribution is equal to the average singlet score, which is 0.31 bits per aligned residue for BLOSUM65 (BLOCKS13+). The corresponding average doublet score, the additional information encoded in inter-site substitution covariation, is around 0.04 bits at modest sequence separations (illustrated in fig.~\ref{fig-MI}). 
Thus, the inter-site substitution correlations carry relatively little information. However, these correlations appear to persist to non-local neighbors, which suggests that the total information from interactions at all sequence separations is substantial. However, figure~\ref{fig-MI} also displays the contributions to this total information from various categories of substitution. The largest contribution, and the only contribution to persist above a sequence separation of 4 residues, represents exactly conserved pairs of residues. This is a rather trivial correlation which is persistent because all parts of two homologous sequences have the same chronological divergence.  All other substitution classes, summing over all sequence separations, contribute no more than 0.1 bits per residue. This is not entirely insignificant, but it is still small compared to the singlet mutual information. Thus non-trivial correlations between substitutions are relatively weak.

\subsection{Homology Detection}

\begin{figure}[tp!]
\begin{center}
\includegraphics{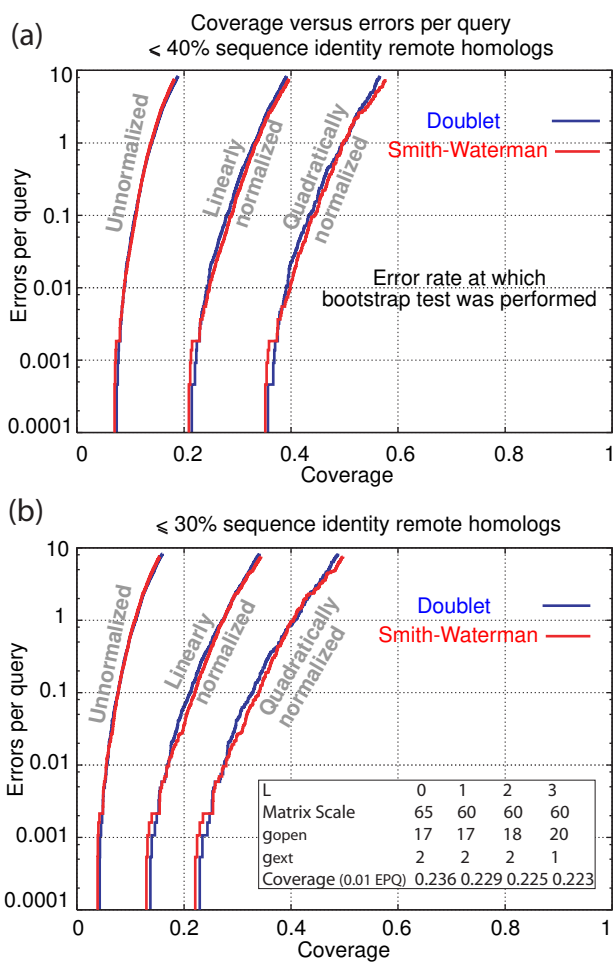}
\caption{These coverage versus
  errors per query plots show that including dipeptide covariation
  information in alignment determination ({\tt doublet}) does not improve
  remote homolog detection. (a) Optimized matrix, gap and
  look-back parameters were used to search the test database with the
  {\tt doublet} and Smith-Waterman algorithms.  This database contains no
  sequence pairs that share more than 40\% sequence identity.  The
  number of correctly identified homologs is shown as a function of
  the number of errors made.  Smith-Waterman outperforms {\tt doublet} over
  all but extremely low error-rates. (b) Remote homolog test using
  only sequence pairs with less than 30\% sequence identity.
  As above, Smith-Waterman correctly identifies more remote homologs
  than the {\tt doublet}
  algorithm.
  Insert: Optimal matrix scale parameter, gap parameters, and corresponding linearly normalized homology detection coverage at 0.01 EPQ, as a function of the covariation distance considered, $L$
  }
\label{CVE}
\end{center}
\end{figure}

The primary use for pairwise alignment methods is to search databases
of previously characterized biological sequences for homologs of the sequence of
interest.  Therefore, the most powerful methods will perform this task
most effectively by assigning true homologs significant statistical
scores and assigning unrelated sequences low statistical scores. Our
assessment methodology compares database search methods on this
criteria.

We compared the {\tt doublet} alignment algorithm against the standard
Smith-Waterman algorithm.  To perform a fair test, we converted raw
scores to statistical scores for both algorithms using the same length
normalized maximum likelihood EVD parameter determination method
\citep{BaileyGribskov-2002-JCompBiol}.
Optimal parameters for gapping, matrix scaling, and distance over which to consider dipeptide correlations
were found using the training database described above.  Then,
the algorithms were evaluated by comparing the relative ability to detect remote homologs within the test dataset, using the parameters optimized on the training dataset.  (Insert, fig.~\ref{CVE}).

The results of a database
search for Smith-Waterman and {\tt doublet}, using only nearest neighboring dipetide covariations,
are shown in Fig.~\ref{CVE}a.
Both the Smith-Waterman and {\tt doublet} methods performed remarkably similarly over all error rates and
normalization schemes.  The linearly normalized coverage at 0.01 errors
per query was slightly higher for Smith-Waterman than {\tt doublet} (Insert, fig.~\ref{CVE}).  From this, we conclude that including dipeptide covariation
information does not improve remote homology detection and, in fact,
slightly degrades performance at this error rate.  We also performed
the same coverage versus errors per query analysis using only
sequences with less than 30\% sequence identity (Fig.~\ref{CVE}b), as it was
previously reported that dipeptide covariation information may be
useful only for detecting these extremely remote evolutionary
relationships  \citep{JungLee-2000-ProteinScience}.  Our results,
however, show that even at this evolutionary distance, dipeptide covariation
scoring does not improve homology detection.

We used Bayesian bootstrap resampling to estimate statistical errors
and to determine if the observed coverage difference was statistically
significant. We found that a 95\% confidence
interval for the coverage difference at 0.01 errors per
query comfortably contained zero difference. Therefore, we cannot distinguish between the remote homolog detection abilities of Smith-Waterman and {\tt doublet}.

We also evaluated the effect of including covariation information over larger sequence separations. As can be seen in table of fig.~\ref{CVE}, incorporating this additional information into alignment scores actually results in a slow degradation of homology detection efficacy.

\section{DISCUSSION}

We have developed, implemented, and tested an alignment algorithm,
{\tt doublet}, that generates the optimal pairwise protein sequence alignment under a scoring
scheme that includes dipeptide covariation information.  Perhaps surprisingly,
and in marked contrast to previous reports, we found that using this
information provides no benefit to remote homolog detection.  The
performance of the {\tt doublet} algorithm for detecting remote homologs is
statistically indistinguishable from the standard Smith-Waterman
algorithm.  

The underlying explanation for this indifference of alignment to dipeptide covariation is that 
substitution correlations are weak on the average (Figs.~\ref{fig-blosum} and \ref{fig-MI}).  Therefore, the average effect of these interactions is insignificant and including covariation in sequence alignment makes very little material difference to remote homology detection. 

We might reasonably question if the training data is at fault. Indeed,
the slight degradation of homology detection as more distant
correlations are included (Insert table, fig.~\ref{CVE}) does indicate that
the doublet substitution matrices contain anomalies, perhaps due to
the training or alignment of the BLOCKS sequences, or perhaps because
of the different sampling of sequences included in BLOCKS compared to those
included in SCOP. The BLOCKS database that we use to train
the doublet substitution matrices contains ungapped
alignments, many of shorter length than the average SCOP protein
domain.   Fikami-kobayashi and co-workers showed that the covariation
signal is strongest within single secondary structure elements
\citep{Fukami-Kobayashi2002}. The poor performance of {\tt doublet},
then, may be due to its applying the 
covariation model too bluntly across entire protein sequences when it
is only applicable within secondary structure elements.
 However, we note that the BLOCKS database
has been used to derive very effective singlet substitution matrices
\citep{Green-2002-ProcIEEE}, and therefore it is implausible
that the substitution signals within the BLOCKS database are
substantially erroneous.  Rather, the observed degradation simply
reinforces the idea that neighboring substitutions are weakly
correlated, particularly when compared to single substitution
correlations, and therefore the doublet signal is readily degraded by
minor anomalies in the data. 

Another line of evidence comes from examining the inter-site amino
acid correlation of single protein sequences \citep{Ycas1958, Weiss2000,
  Crooks2004a, Crooks2004b}. Neighboring amino acids are almost
entirely uncorrelated; the nearest neighbor mutual information has been
estimate as only 0.006 bits \citep{Crooks2004a}. This lack of
sequence correlation is consistent with (but does not require) small
inter-site substitution correlations.

In should be emphasized, however, that the observation of weak average
dipeptide covariation does not negate the possibility of strong,
interesting covariation in particular instances, such as
CP$\leftrightarrow$Cx, or within particular families. Moreover, it is
conceivable that covariation information could be used more
judiciously, thereby improving alignment results.  For example, as previously discussed, one
might include doublet-type scoring information only for residue pairs
that are likely to be within the same secondary structural
element. Similarly, one might examine the covariation of residues that
are proximate in the tertiary structure, rather than along the sequence
\citep{Rodionov1994, Lin2003}. 
However, residues that are proximate in space are also only weakly correlated\citep{Cline2002,Crooks2004b}, and the
inter-residue mutual information is not improved by foreknowledge of
the local structure environment \citep{Crooks2004a,Crooks2004b}.
Therefore, we suspect that such approaches will also not have dramatic
effects on protein sequence alignment.

In conclusion, the ubiquitous assumption that neighboring sites along a protein sequence evolve independently appears to be  generally appropriate. This leads to fast, elegant and effective algorithms for protein sequence alignment and homology detection.

\subsection*{Acknowledgments}
Author Contributions:  R.E.G. and G.E.C. jointly conceived and designed the {\tt doublet} alignment algorithm and co-wrote this paper, with guidance from S.E.B.; G.E.C. was responsible for creating the doublet BLOSUM substitution matrices and R.E.G. for the statistical comparison of {\tt doublet} to Smith-Waterman.
We would like to thank Emma Hill,  Sandrine Dudoit and Jeff Thorne for helpful discussions and suggestions.  This work was supported by the National Institutes of Health 
(1-K22-HG00056) and an IBM Shared University Research grant.  G.E.C. received funding from the Sloan/DOE postdoctoral fellowship in computational molecular biology. S.E.B. is a Searle Scholar (1-L-110).


\section*{APPENDIX: Estimating probabilities from counts with a prior of uncertain reliability. }

A common problem is that of estimating a discrete probability distribution, 
$\theta = \{\theta_1, \theta_2, \ldots ,\theta_k\}$,  given a limited number of samples drawn from that distribution, 
summarized by the count vector $n = \{n_1, n_2,\ldots,n_k\}$, and a reasonable 
{\it a priori} best guess for the distribution $\theta\approx\pi = \{\pi_1,\pi_2,\ldots,\pi_k\}$. 
(For a general introduction, see \citealt{DurbinEddy1998}.) 
%
%
This guess may simple be the uniform probability, $\pi_i = 1/k$, which amounts to asserting that, as far as we know, all possible observations are equally likely. At other times, we may know some some more detailed approximation to the distribution $\theta$. 

For example, in the present case we wish to estimate the probabilities of substituting a pair of amino acid residues by another residue pair, given the number of times that this substitution has been observed in the training dataset. This probability is hard to estimate reliably since the distribution is very large with $20^4=160,000$ dimensions. Moreover, many of the possible observations occur very rarely.
However, substitutions at different sites are not strongly correlated, and therefore we may approximate the doublet substitution probabilities by a product of single substitution probabilities. Since the dimensions of these marginals are relatively small we can accurately estimate them from the available data, and thereby construct a reliable and reasonable initial guess for the full doublet substitution distribution.

In the common and conventional pseudocount approach, we assume that the 
distribution $\pi$ was estimated from  $A$ 
previous observations. These pseudocounts, $\alpha_ i = \pi_i A$, are then proportionally averaged with the real observations ($N = \sum_i n_i$) to provide an estimate of $\theta$;
\begin{equation}
 \theta_i = \frac{  \alpha_i+n_i}{A + N} .
\end{equation}
This prescription is intuitively appealing. When the total number of real counts is much less than the number of pseudocounts ($N \ll A$) the prior dominates, and the estimated distribution is determined by our initial guess, $\theta \approx \pi$.  In the alternative limit that the real observations greatly outnumber the pseudocounts ($N\gg A$) the estimated distribution is given by the frequencies $\theta_i = n_i/ N$.
However, it is not immediately obvious how to select $A$, although
many heuristics have been proposed, including $A= 1$, $A=k$ (Laplace), and $A=\sqrt {N}$  \citep[e.g. ][]{Lawrence1993,DurbinEddy1998,Nemenman2001}.    
Essentially, this total pseudocount parameter represents our confidence that the initial guess $\theta\approx\pi$ is accurate, since the larger the total pseudocount the more data is required to overcome this assumption.

Within a Bayesian approach we can avoid this indeterminacy by admitting that, {\it a priori}, we do not know how confidant we are that $\pi$ approximates $\theta$. 
The probability $P(n|\theta)$ of independently sampling a particular set of observations, $n$, given the underlying sampling probability, $\theta$,  follows the multinomial distribution, the multivariate generalization of the binomial distribution;
\begin{equation}
{\mathcal M}(n| \theta) =    \frac{1}{M(n)}  \prod_{i=1}^{k} \theta_{i}^{n_i},
\quad M(n) = \frac{ \prod_i n_i !}{(\sum_i n_i)!} .
\end{equation}

The prior probability of the sampling distribution $P(\theta)$ is typically modeled with a Dirichlet distribution,
\begin{equation}
{\mathcal D}(\theta | \alpha) = \frac{1}{Z(\alpha) }
\prod_{i=1}^{k} \theta_i^{(\alpha_i-1)},
\quad Z(\alpha) =  \frac{\prod_i \Gamma(\alpha_i)}{\Gamma(A)} .
\end{equation}
\noindent where $\sum_i \theta =1 $, $\alpha_i>0$ and $A= \sum_i \alpha_i$. Note that the mean of a Dirichlet is 
\begin{equation}
\mathrm E [\theta_i] = \frac{\alpha_i}{A} .
\label{mean_theta}
\end{equation}
\noindent Therefore, we may fix the parameters of the Dirichlet prior by equating our initial guess, $\pi$, with the mean prior distribution: $\pi = \alpha/A$.  If we can fix the scale factor $A$, then we can combine the prior and observations using Bayes' theorem.
\begin{equation}
   P(\theta | n) = \frac{P(n | \theta) P(\theta) }{P(n) } \, .
   \label{bayes}
\end{equation}

Because the multinomial and Dirichlet distributions are naturally conjugate, the posterior distribution $P(\theta|n)$ is also Dirichlet.
\begin{eqnarray}
P(\theta|n) &\propto& {\mathcal M}(n| \theta) {\mathcal D}(\theta | A\pi )  \nonumber \\
		 &\propto&  \prod_{i=1}^{k} \theta_{i}^{(A\pi_i + n_i-1)}, \nonumber \\
                    &=& {\mathcal D}(\theta | A\pi + n) 
\end{eqnarray}
The last line follows because the product in the previous line is an unnormalized Dirichlet with parameters $(A\pi +n)$, yet the probability $P(\theta|n)$ must be correctly normalized.

Given multinomial sampling and a Dirichlet prior, the probability of the data is given by the under-appreciated multivariant negative hypergeometric distribution  \citep[][Eq. 11.23]{JohnsonKotz-1969, DurbinEddy1998};
\begin{eqnarray}
P(n) 
&=& \int d\theta \; P(n |\theta) P(\theta) 
,\nonumber\\ &=& 
\int d\theta \; {\mathcal M}(n|\theta)  {\mathcal D}(\theta | A\pi ) ,
\nonumber\\
	    &=&  \frac{1}{Z(A\pi )}\frac{1}{M(n)} \int d\theta 
	    \prod_{i=1}^{20} \theta_i^{(A\pi_i + n_i -1)}  ,
	    \nonumber\\
	    &=&  \frac{ Z(A\pi +n) }{Z(A\pi )M(n)}
	     \equiv {\mathcal H'}( n | A\pi +n ) .
\label{hypergeometric}
\end{eqnarray}
Again, the last line follows because the product in the previous line is an unnormalized Dirichlet with parameters $(A\pi + n)$. Therefore, the integral over $\theta$ must be equal to the corresponding Dirichlet normalization constant,  $Z(A\pi + n)$. 
Note that, confusingly, the negative hypergeometric distribution is sometimes called the inverse hypergeometric, an entirely different distribution, and vice versa.

Since we do know a reasonable value for the scale factor $A$ we cannot use a simple Dirichlet prior. As an alternative, we explicitly acknowledge our uncertainly about $A$ by building this indeterminacy into the prior itself. Rather than a single Dirichlet, we use the Dirichlet mixture; 
\begin{equation}
P(\theta|\pi) = \int_0^{\infty} dA \;  {\mathcal D}(\theta | A\pi ) P(A) .
\end{equation}
\noindent
The distribution $P(A)$ is a hyperprior, a prior distribution placed upon a parameter of the Dirichlet prior.  Following the same mathematics as Eqs.~\ref{bayes}-\ref{hypergeometric}, we find that the posterior distribution is the Dirichlet mixture
\begin{eqnarray}
P(\theta|n) &=& \int_0^{\infty} dA \; {\mathcal D}(\theta | A\pi  + n) P(A|n)  \, ,
\label{dirichlet_mixture}
\end{eqnarray}
\noindent where
\begin{eqnarray}
P(A|n) &=&\frac{P(A) {\mathcal H'}( n |  A\pi +n ) }
{\int_0^{\infty} dA  \; P(A) {\mathcal H'}( n | A\pi + n )  } \, . 
\end{eqnarray}

\begin{figure}[t]
\begin{center}
\includegraphics{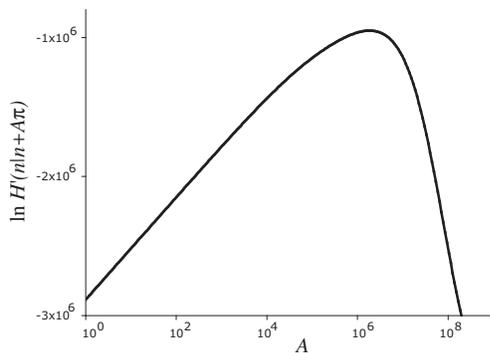}
\caption{ The likelihood of observations as a function of the scale parameter $A$. With multinomial sampling and a Dirichlet prior the likelihood of the data follows the negative hypergeometric distribution, $H'(n | A\pi+n )$, where $n$ is the count vector of observations, $\pi$ is the mean prior estimate of the sampling distribution, and $A$ is a scale parameter (Eq.~\ref{hypergeometric}). Given a large number of observations (here, $N = \sum n_i$ is about $10^7$) the probability of the data is a smooth and very sharply peaked function of the scale parameter $A$.
}
\label{fig-hypergeometric}
\end{center}
\end{figure}

In principle, we have to select and parameterize a functional form for the hyperprior, $P(A)$.
For example, an exponential distribution, $P(A) =\lambda \exp(-\lambda A)$, with mean $1/\lambda$, might be appropriate.
Fortunately, we can often avoid selecting an explicit hyperprior.
In practice, given sufficient data, the probability of that data $P(n|A)$ is a smooth, sharply peaked function of $A$. This is illustrated in figure~\ref{fig-hypergeometric} using $10^7$ observations of the 160,000 dimensional doublet substitution probability, where the mean prior distribution is taken to be the product of singlet substitutions probabilities%
.  If the prior distribution of $A$ is reasonable, and neither very large nor very small over the range of interest, then the posterior distribution $P(A|n)$ will also be very strongly peaked. Moreover, the location of that peak will be almost totally independent of the prior placed on $A$. In this limit the posterior Dirichlet mixture (Eq.~\ref{dirichlet_mixture}) reduces to the single component that maximizes the probability of the data;
\begin{eqnarray}
P(\theta|n) &\approx&  {\mathcal D}(\theta | A\pi  + n), \nonumber \\
A &=& \mathrm{argmax}_{A} P(A|n)  \approx \mathrm{argmax}_{A} P(n| A) ,\nonumber \\
P(n|A) &=& {\mathcal H'}( n | A\pi+n ) .
\end{eqnarray}
\noindent Here, $\mathrm{argmax}_x f(x)$ is the value of $x$ that maximizes that function $f(x)$.

Given any function of $\theta$, the average of the function across the posterior distribution (the posterior mean estimate (PME) or Bayes' Estimate) minimizes the mean squared error of that estimate. In particular, the posterior mean estimate of $\theta$ (Eq.~\ref{mean_theta}) is
\begin{equation}
\theta^{\mathrm{PME}}_i = \frac{A\pi_i  + n_i}{A + N} .
\end{equation}

Taken altogether, our practice is to take the raw doublet substitution counts and construct a mean prior distribution $\pi$ based upon the approximation that substitutions on neighboring sites are uncorrelated
. We then find the scaling factor $A$ that maximizes the negative hypergeometric probability ${\mathcal H'}( n | A\pi +n )$.  For our data the total number of observations $N$ is around $10^7$, for which the optimal scale factor $A$ was found to be about $10^6$%
. The posterior mean estimate of the doublet substitution distribution 
is then used to construct the doublet substitution matrix%
. 
Code for constructing doublet substitution matrices using this procedure and for finding the optimal prior and posterior, given any set of observations and $\pi$, a best guess for the true distribution $\theta$, is available from our web site (\url{http://compbio.berkeley.edu}), along with other code and data for this work. Our programs make extensive use of the Open Sourced GNU Scientific Library (GSL)  \citep{GSL, MersenneTwister}.


 \bibliographystyle{bioinformatics} 
\bibliography{doublet}

\end{document}